\documentclass[prx,twocolumn,aps]{revtex4-1}
\usepackage{epsfig, amsmath}
\usepackage{bm}
\usepackage{hyperref}
\newcommand{\B}[1]{{\bm{#1}}}
\newcommand{\C}[1]{{\mathcal{#1}}}
\usepackage[latin1]{inputenc}
\newcommand{\beq}{\begin{equation}}
\newcommand{\eeq}{\end{equation}}
\newcommand{\bea}{\begin{eqnarray}}
\newcommand{\eea}{\end{eqnarray}}
\newcommand{\pa}{\partial}
\newcommand{\tr}{\hbox{tr}}

\begin{document}
\title{Notch fracture toughness of glasses: Rate, age and geometry dependence}

\author{Manish Vasoya$^{1}$, Chris H. Rycroft$^{2,3}$ and Eran Bouchbinder$^1$}
\affiliation{$^1$ Chemical Physics Department, Weizmann Institute of Science,
Rehovot 7610001, Israel\\
$^2$ Paulson School of Engineering and Applied Sciences, Harvard University,
Cambridge, MA 02138, United States\\
$^3$ Department of Mathematics, Lawrence Berkeley Laboratory, Berkeley, California 94720, United States}


\begin{abstract}
Understanding the fracture toughness (resistance) of glasses is a fundamental
problem of prime theoretical and practical importance. Here we theoretically
study its dependence on the loading rate, the age (history) of the glass and the
notch radius $\rho$. Reduced-dimensionality analysis suggests that the notch
fracture toughness results from a competition between the initial, age- and
history-dependent, plastic relaxation timescale $\tau^{pl}_0$ and an effective
loading timescale $\tau^{ext}(\dot{K}_I,\rho)$, where $\dot{K}_I$ is the tensile
stress-intensity-factor rate. The toughness is predicted to scale with
$\sqrt{\rho}$ independently of $\xi\!\equiv\!\tau^{ext}\!/\tau^{pl}_0$ for
$\xi\!\ll\! 1$, to scale as $T\sqrt{\rho}\,\log(\xi)$ for $\xi\!\gg\!1$ (related
to thermal activation, where $T$ is the temperature) and to feature a
non-monotonic behavior in the crossover region $\xi\!\sim\!{\cal O}(1)$ (related
to plastic yielding dynamics). These predictions are verified using novel 2D
computations, providing a unified picture of the notch fracture toughness of
glasses. The theory highlights the importance of timescales competition and far
from steady-state elasto-viscoplastic dynamics for understanding the toughness,
and shows that the latter varies quite significantly with the glass age
(history) and applied loading rate. Experimental support for bulk metallic
glasses is presented.
\end{abstract}
\maketitle

\section{Introduction}
\label{sec:intro}

The fracture toughness, i.e. the ability to resist failure in the presence of a
crack, is a basic physical property of materials~\cite{Meyers_Book}. From a practical perspective, this property is a
major limiting factor in the structural integrity of a broad range of systems. From
a theoretical perspective, the fracture toughness challenges our understanding
of the strongly nonlinear and dissipative response of materials under extreme
conditions, approaching catastrophic failure. Consequently, obtaining a basic
understanding of the fracture toughness of materials is a fundamentally important
problem.

Quantitatively predicting the fracture toughness of glassy materials, which lack
long-range crystalline order and are characterized by intrinsic disorder, is a
particularly pressing problem. Glassy materials exhibit
unique and intriguing physical properties as compared to their crystalline
counterparts~\cite{GI75, SHR07, C08, Review2010, CM11, W12, EID13, HSM16}. For example, glassy solids typically exhibit a strength and an elastic
limit that significantly exceed those of crystalline alloys of similar
composition due to the absence
of mobile dislocation defects. Instead, glassy solids deform irreversibly by
immobile and localized structural rearrangements~\cite{SP77,AS79,FL98}, at sites termed
Shear-Transformation-Zones (STZ), which are not yet fully understood.

Glassy materials are intrinsically out-of-equilibrium, hence their physical
properties depend on their history and age (see, for example,~\cite{S78, EAN96}). Moreover, these materials typically
do not feature strain-hardening, i.e.~an increase in the deformation
resistance with increasing deformation, which is commonly observed in
crystalline alloys~\cite{GI75, SHR07, C08, Review2010, CM11, W12, EID13, HSM16}. Finally, glassy materials also feature rate effects that are
far from understood~\cite{GI75, SHR07, C08, Review2010, CM11, W12, EID13, HSM16}.

In the last few decades significant progress has been made in using conventional
casting techniques to obtain multi-component amorphous alloys in bulk forms, the
so-called Bulk Metallic Glasses (BMG)~\cite{G95, I00, J02, L03, WDS04, AG06}. The emergence of this new
family of glasses, which possess superior properties, has triggered intense
research activity and holds a great promise for a wide range of engineering
applications~\cite{GI75, SHR07, C08, Review2010, CM11, W12, EID13, HSM16}. The fracture toughness of these materials, though, still raises
serious concerns and hence the current usage of BMG in structural applications
is limited.

Consequently, the fracture toughness of glasses, and of BMG in particular, has
been the subject of various recent studies~\cite{01Lewandowski, 05LWG, Notch, damage-tolerant, FD01_1, FONYKI07, LSN06, TNR07, XRM10, HA09, TRN09, NSPTR09, RB12, NTSNR15, Sun2015, LGB15, DLZWXG15, FD01, BBPBPR08, JLMD08, Lo2008R, Lo2008, MGZNLG11, An2011, Guan2013, Singh2014, Maas2015}. Yet, there is no
complete understanding of the resistance of glassy materials to catastrophic
failure in the presence of a notch defect --- the notch fracture toughness --
and its dependence on various physical parameters. Our goal in this paper is to
offer a comprehensive theoretical picture of the dependence of the notch
fracture toughness of glasses on the loading rate, the glass age/history, the
notch radius and the temperature (below the glass transition temperature).

Our main result, obtained through a reduced-dimensionality theoretical analysis
and extensive 2D spatiotemporal computations based on a novel numerical
method~\cite{RSB15}, highlights the essential role played by timescales competition
in determining the notch fracture toughness of glassy materials. The competing
timescales involve an effective applied loading timescale (depending on the notch radius
of curvature, and on the global geometry and loading rate) and the initial,
age-dependent, plastic (dissipative) relaxation timescale (depending on the
glass age and history)~\cite{comment}. Once properly identified, the ratio of the two
timescales
is shown to control the fracture toughness over a wide range of
physical conditions. The dependence of the toughness on the dimensionless
timescale ratio is quantitatively derived and is shown to feature a
{\em non-monotonic} behavior. Our results are shown to be consistent with
existing experimental data and offer various new predictions.

\section{Problem formulation}
\label{sec:formulation}

The fracture toughness quantifies the amount of dissipation involved in crack
propagation. Consequently, one needs to account for the irreversible deformation
of the material, and its interplay with reversible (elastic) deformation.
Inertia plays little role in fracture {\em initiation} under a wide range of
conditions and standard engineering testing protocols, hence we focus on
quasi-static stress equilibrium described by
\begin{eqnarray}
\nabla\!\cdot\B\sigma = \B 0 \ ,
\end{eqnarray}
where $\B \sigma$ is the Cauchy stress tensor. We consider then a general
hypo-elasto-viscoplastic material described by
\begin{eqnarray}
\B D^{tot}(\B v) = \B D^{el}(\B\sigma, \B v) + \B D^{pl}(\B\sigma,...)\ .
\end{eqnarray}
Here $\B D^{tot}\!=\!\case{1}{2}[\nabla {\B
v} \!+\! \left( \nabla {\B v}\right)^T]$ is the total rate of deformation
tensor, where $\B v(\B r,t)$ is the Eulerian
velocity field and $(\B r, t)$ are the spatiotemporal coordinates. $\B
D^{el}\!=\!\pa_t \B \epsilon+\B v \cdot \nabla \B
\epsilon + \B \epsilon \cdot \B \omega -  \B \omega \cdot \B \epsilon$
is the elastic
rate of deformation tensor, where $\B \omega\!=\! \case{1}{2}[\nabla {\B v}
\!-\! \left( \nabla {\B
v}\right)^T]$ is the spin tensor and the strain tensor $\B\epsilon$ is related
to $\B \sigma$ through Hooke's law $\B \sigma \!=\!
K\,\tr\B\epsilon\,\B 1+2\mu\left(\B\epsilon-\case{1}{3}\tr\B \epsilon\,\B
1\right)$. $K$ and $\mu$ are the bulk and shear moduli, respectively. $\B
D^{pl}(\B\sigma,...)$ is the
plastic rate of deformation tensor, which encapsulates the relevant
physics of the dissipative deformation of glasses. The ellipsis stands for
additional dependencies,
e.g. on the temperature and on structural internal state variables.

We adopt the non-equilibrium thermodynamic Shear-Transformation-Zones (STZ)
model, as in~\cite{RB12}, where
\begin{eqnarray}
\label{eq:simple_STZ}
\B D^{pl}(\B s, T, \chi) &=&  e^{-\tfrac{e_z}{k_B\chi}}\,\frac{\C
C(\bar{s},T)}{\tau}\,\left[1 - \frac{s_y}{\bar{s}} \right] \frac{\B s}{\bar{s}}\
,\\
\label{eq:chi}
c_0\,\dot\chi &=& \frac{{\B D}^{pl}\!\!:\!{\B s}}{s_y}\,\,(\chi_\infty - \chi) \
,
\end{eqnarray}
for $\bar{s}\!\ge\! s_y$ and $\B D^{pl}\!=\!0$ otherwise. This model, despite
its relative simplicity, has been shown to capture a wide range of driven glassy
phenomena~\cite{FL98, RB12, L04, BLP-07-I, BLP-07-II, MLC07, LA08, BL09_1, BL09_2, BL09_3, Manning2009, FL11, BL11, RG12, PB14, Kamrin2014}. $\B s\!=\!\B\sigma \!-\! \tfrac{1}{3}\tr\B\sigma \,\B 1$
in Eqs.~\eqref{eq:simple_STZ}-\eqref{eq:chi} is the deviatoric stress tensor,
its magnitude is
$\bar{s}\!\equiv\!\sqrt{{\B s}\!:\!{\B s}/2}$, and $s_y$ is the shear yield
stress. $\chi$ is an effective disorder temperature which quantifies the
intrinsic structural state of the glass~\cite{L04, BL09_2}, $e_z/k_B$ is a typical STZ formation
energy over Boltzmann's constant,
$\C C(\bar{s},T)/\tau$ is the rate at which STZ make transitions between their
internal states. $\tau^{-1}$ is a molecular vibration rate and $T$ is the bath
temperature, assumed to be well below the glass temperature (such that
spontaneous aging is neglected in Eq.~\eqref{eq:chi}). $c_0$ is an
effective dimensionless heat capacity and $\chi_\infty$ is the steady state
value of $\chi$.

The STZ transition rate is taken to be of the form
\begin{eqnarray}
\label{Cal_C}
\C C(\bar{s},T) \!=\! \left\{\begin{array}{ll}
\!\!e^{-\tfrac{\Delta}{k_B T}}
\cosh\left[\tfrac{\Omega\,\epsilon_0\,\bar{s}}{k_B T}\right]
&\quad\text{for}\quad\  \Omega\,\epsilon_0\bar{s} < \Delta \vspace{0.1cm}\\
           \!\!\tfrac{\Omega\,\epsilon_0\,\bar{s}}{2\Delta}
&\quad\text{for}\quad \  \Omega\,\epsilon_0\bar{s} \ge \Delta \ .
\end{array}\right.
\end{eqnarray}
It corresponds to a linearly stress-biased thermal activation process at
relatively small stresses, where $\Delta$ is the typical energy activation
barrier, $\Omega$ is the typical activation volume and $\epsilon_0$ is the
typical
local STZ strain. In the presence of the high stresses near a tip of a crack,
$\Omega\,\epsilon_0\bar{s}$ may become larger than $\Delta$, in which case we
assume that the exponential thermal activation form crosses over to a much weaker
dependence associated with a linear, non-activated, dissipative mechanism~\cite{LA08}.
As $\Delta\!\gg\!k_B T$, the two regimes connect continuously, but not
differentiably. This crossover in the form of the STZ transition rates, from
exponential thermal activation to a much weaker athermal power-law (here a
linear relation, which allows for analytic progress), will turn out below
to have important implications for the toughness.

\begin{figure}[h]
\centerline{\includegraphics[width=0.49\textwidth]{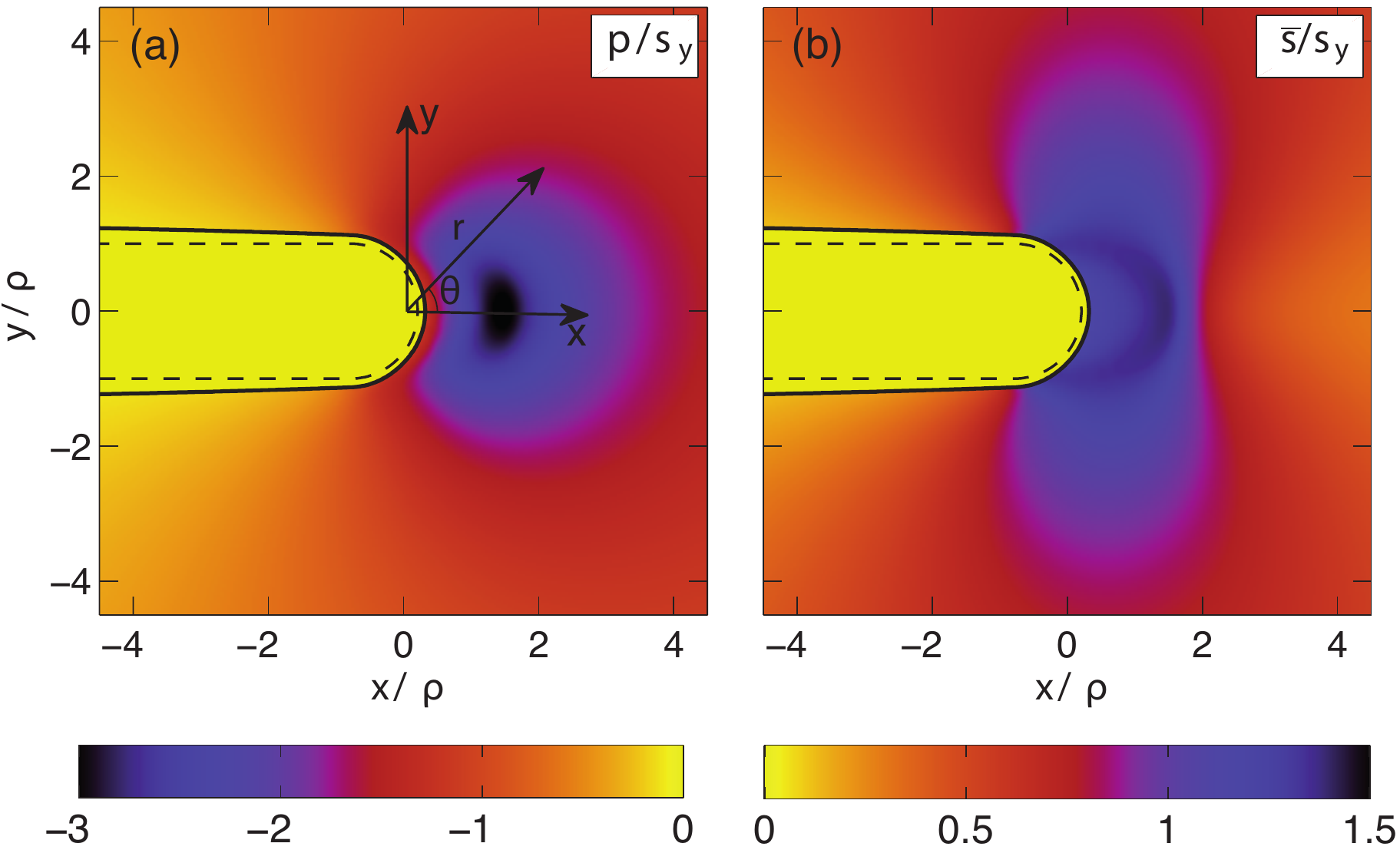}}
\caption{(Color online) The problem setting and an example of a numerical solution in the near
notch root region. (a) The hydrostatic pressure and (b) the magnitude of the
deviatoric stress, both normalized by the shear yield stress $s_y$, are
shown. The dashed-dotted line corresponds to the initial notch state and the
solid line to a deformed state with $K_I\!=\!30~\rm{MPa\sqrt{m}}$. A small portion of the
simulation domain $-20\!\le\!x/\rho,~y/\rho\!\le\!20$, near the notch root, is shown. A fixed coordinate system located a distance $\rho/5$ behind the initial notch root, with both Cartesian $(x,y)$ and polar $(r,\theta)$ coordinates, is shown on panel (a). The calculation was done using a $1025\!\times\!1025$ grid.}
\label{fig1}
\end{figure}

This elasto-viscoplasticity model is used to formulate a plane-strain fracture problem where traction-free boundary conditions are imposed on a blunted straight notch (crack) with an initial root radius $\rho$
(cf. Fig.~\ref{fig1}) and the universal linear elastic mode I (tensile) crack tip velocity
fields~\cite{BR99}
\begin{eqnarray}
\label{Irwin}
{\B v}(r,\theta,t) = \frac{\dot{K}_I(t)}{\mu}\sqrt{\frac{r}{2\pi}}\,{\B
F}(\theta) \quad\hbox{for}\quad r\gg\rho
\end{eqnarray}
are imposed on a scale much larger than $\rho$. Here $\dot{K}_I$ is the mode I
stress-intensity-factor rate, which measures the intensity of the linear elastic
singularity $\nabla {\B v}\!\sim\!1/\sqrt{2\pi
r}$ at $\rho\!\ll\!r\!\ll\!L$, where $L$ is a macroscopic lengthscale in the
global fracture problem (e.g. the sample size). $(r,\theta)$ is a polar
coordinate system whose origin is set a distance $\rho/5$ behind the notch root,
$\theta\!=\!0$ is the symmetry axis and ${\B F}(\theta)$ is a known universal
function~\cite{BR99, RB12}. In such a boundary layer formulation, the
stress-intensity-factor uniquely couples the inner scales near the notch root to the outer scales, and hence can be
controlled independently without solving the global fracture problem~\cite{BR99}.

The fracture toughness is the critical value of the stress-intensity-factor,
$K_{I\!c}$, at which crack propagation initiates and global failure occurs.
Recent work~\cite{RB12, FD01, BBPBPR08, JLMD08, Lo2008R, Lo2008, MGZNLG11, An2011, Guan2013, Singh2014, Maas2015} suggests that this onset (and in fact also the subsequent
propagation~\cite{RB12}) is controlled by a local cavitation instability occurring
when the hydrostatic tension $\case{1}{3}\tr\B\sigma$ surpasses a threshold
level $\sigma_c$. We adopt this failure criterion here.

While a large part of the analysis below is performed in terms of dimensionless
parameters, we nevertheless
consider realistic material parameters corresponding to Vitreloy 1, a widely
studied BMG, identical to those reported in~\cite{RB12}. That is, we
use:
$T\!=\!400$~K, $e_z/k_B\!=\!21000$~K, $s_y\!=\!0.85$~GPa, $\mu\!=\!37$~GPa,
$K\!=\!122$~GPa, $\tau\!=\!10^{-13}$~s, $\epsilon_0\!=\!0.3$,
$\Omega\!=\!300$~{\AA}$^3$, $c_0\!=\!0.4$, $\Delta/k_B\!=\!8000$~K and
$\chi_{\infty}\!=\!900$~K.
For the initial conditions we use $\B\sigma(\B r,t\!=\!0)\!=\!0$ and $\chi(\B
r,t\!=\!0)\!=\!\chi_0$,
where $\chi_0$ describes the initial structural state of the glass which depends
on its history. For example, it may be affected by the cooling rate at which the glass has been
formed, annealing and other heat treatments, aging time and
pervious deformation. The model's setup is shown in Fig.~\ref{fig1}.

\section{Theory and analysis}
\label{sec:analysis}

Our major goal is to study the dependence of the fracture toughness $K_{I\!c}$
on the structural state of the glass $\chi_0$, on the stress-intensity-factor
rate $\dot{K}_I$, on the notch radius of curvature $\rho$ and on the temperature
$T$ below the glass transition temperature. We address the problem of
calculating $K_{I\!c}(\chi_0, \dot{K}_I, \rho, T)$ by a reduced-dimensionality
theoretical analysis and 2D numerical computations. The latter, an example of
which is shown in Fig.~\ref{fig1}, are based on a recently developed numerical
method that can handle physically realistic loading rates, which is essential
for understanding the properties of the toughness. Preliminary numerical results
addressing this problem appeared in~\cite{RB12}.

To gain some analytic insight into the fracture toughness, we perform a
reduced-dimensionality analysis which aims at describing the behavior of a
representative material element near the notch root. We further simplify the
problem by eliminating its tensorial nature, focusing only on the magnitude of
the deviatoric component of the relevant tensors. In particular, we neglect
altogether the hydrostatic part of the stress tensor $\B \sigma$ and replace its
deviatoric part $\B s(\B r,t)$ by a space-independent scalar $s(t)$, and
$\chi(\B r,t)$ by $\chi(t)$. Similarly, we replace the space-dependent elastic
and plastic rate of deformation tensors in the problem by their
space-independent scalar counterparts ${\B D}^{el}(\B r,t) \!\to\!
\dot{s}(t)/\mu$ and ${\B D}^{pl}(\B r,t) \!\to\! D^{pl}(t)$, with
$D^{pl}(s,\chi)\!=\!\tau^{-1} e^{-\frac{e_z}{k_B\chi}} {\C C}(s, T)(1-s_y/s)$.

The crucial last step is to relate the global loading and geometry of the
system, captured by the stress-intensity-factor $K_I$, and the effective total
rate of the deformation near the notch root, taking into account both the strong
stress amplification associated with the linear elastic square root singularity
and the characteristic lengthscale inherited from the notch radius of curvature.
A natural way to do this is through the replacement
\begin{equation}
{\B D}^{tot}(\B r,t) \,\to\, \frac{\dot{K}_I}{\mu \sqrt{2\pi\rho}} \ .
\end{equation}
With these replacements, Eqs.~\eqref{eq:simple_STZ}-\eqref{eq:chi} transform
into a set of coupled nonlinear ordinary differential equations
\begin{eqnarray}
\label{eq:dot_s}
\dot{s} &=& \frac{\dot{K}_I}{\sqrt{2\pi\rho}} - \mu\,D^{pl}(s,\chi) \ ,\\
\label{eq:dot_chi}
c_0\dot{\chi} &=& \frac{D^{pl}(s,\chi)\,s}{s_y}\,(\chi_\infty - \chi) \ .
\end{eqnarray}

Obviously, Eqs.~\eqref{eq:dot_s}-\eqref{eq:dot_chi} miss many features of the
full 2+1 dimensional spatiotemporal dynamics of the problem, such as the
tensorial nature of the basic quantities, the coupling between the deviatoric
and hydrostatic parts of the deformation/stress, the time evolution of the
radius
of curvature $\rho(t)$ and the propagation of yielding fronts in the notch root
region.
Yet, as will be shown below, they capture important aspects of the fracture
toughness. The first step in analyzing
Eqs.~\eqref{eq:dot_s}-\eqref{eq:dot_chi} is to identify a proper set of
dimensionless physical parameters that control their behavior. In this context,
we stress that elasto-viscoplasticity is intrinsically linked to a competition
between different timescales. Moreover, glassy response is sensitive to the {\em
initial} structural state of the material (affected by its age, cooling rate,
previous deformation, etc.), which must play a crucial role in far from
steady-state physical properties such as the fracture toughness.

To capture this timescale competition, we define an {\em initial} plastic
relaxation timescale (inverse rate) as
$\tau^{pl}_0(\chi_0)\!\equiv\!\tau\,e^{\frac{e_z}{k_B\chi_0}}$, an effective applied timescale (again, an inverse rate) as
$\tau^{ext}(\dot{K}_I,\rho)\!\equiv\!\mu\sqrt{2\pi\rho}/\dot{K}_I$ and their
ratio
\begin{equation}
\xi(\chi_0, \dot{K}_I, \rho) \equiv \frac{\tau^{ext}}{\tau_0^{pl}} =
\frac{\mu\sqrt{2\pi\rho}}{\tau\dot{K}_I}e^{-\frac{e_z}{k_B\chi_0}} \ .
\end{equation}
This dimensionless quantity plays a central role in what follows. It is
important to note that $1/\tau^{ext}$ is {\em not} the externally applied
strain-rate, but rather the effective strain-rate experienced by the near notch
region.
The effective strain-rate at the innermost scale $r\!\simeq\!\rho$ is
significantly amplified relative to the externally applied strain-rate,
characterizing the outermost scale $L$, according to the linear elastic square
root singularity.

We also define $\tilde{e}_z\!\equiv\!e_z/k_B\chi_0$,
$\tilde{\chi}_\infty\!\equiv\!\chi_\infty/\chi_0$,
$\tilde{\mu}\!\equiv\!\mu c_0 /s_y$, $\tilde{s}\!\equiv\!s/s_y$,
$\tilde{\chi}\!\equiv\!\chi/\chi_0$ and
$\tilde{t}\!\equiv\!t\dot{K}_I/s_y\sqrt{2\pi\rho}$. In terms of these
dimensionless quantities, Eqs.~\eqref{eq:dot_s}-\eqref{eq:dot_chi} take the form
\begin{eqnarray}
\label{eq:dimless_s}
\dot{\tilde{s}} &=& 1 - \xi f(\tilde{s},\tilde\chi)\ ,\\
\label{eq:dimless_chi}
\tilde\mu\, \dot{\tilde\chi} &=& \xi
f(\tilde{s},\tilde\chi)\,\tilde{s}\,(\tilde\chi_\infty\!-\!\tilde\chi) \ ,
\end{eqnarray}
with $f(\tilde{s},\tilde\chi)\!\equiv\!e^{\tilde{e}_z(1-\tilde\chi^{-1})}{\C
C}(s,T)(1-\tilde{s}^{-1})$ for $\tilde{s}\!\ge\!1$ (for $\tilde{s}\!<\!1$ we
have $f\!=\!0$). It should be noted that nondimensionalizing differential
equations using an initial condition, in our case $\chi_0$, might appear
unnatural. Yet, it is a choice that is dictated by the physics of glasses, which
exhibit a rather unique dependence on the initial state.

To proceed, we distinguish between two regimes, one in which the deviatoric
stress significantly surpasses $\Delta/\Omega\,\epsilon_0$, where ${\C
C}\!\sim\!s$, and one in which the deviatoric stress remains close to
$\Delta/\Omega\,\epsilon_0$, where ${\C C}$ varies exponentially with the stress
(cf. Eq.~\eqref{Cal_C}). We focus first on the former and for the sake of
simplicity set $2\Delta/\Omega\,\epsilon_0\!=\!s_y$, which means that we exclude
thermal activation altogether in this part of the reduced-dimensionality
analysis.
\begin{figure}[h]
\centerline{\includegraphics[width=0.48\textwidth]{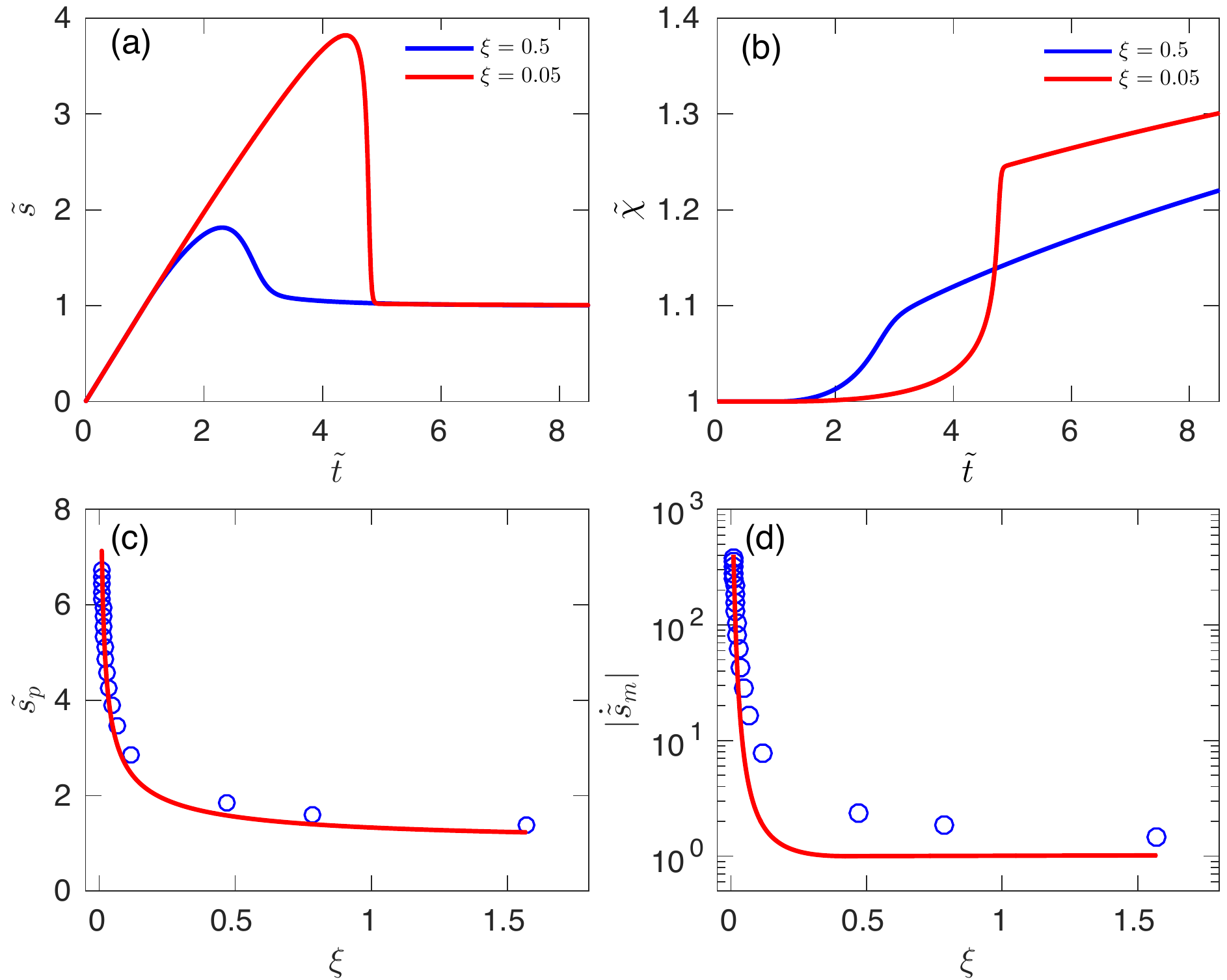}}
\caption{(Color online) The solution of Eqs.~\eqref{eq:dimless_s}-\eqref{eq:dimless_chi}, for two values of $\xi$ (separated by an order of magnitude). The stress is shown in panel (a) and the effective temperature in panel (b). We used $\tilde{\mu}\!=\!15.07$,
$\tilde{e}_z\!=\!35$, $\tilde{\chi}_{\infty}\!=\!1.5$ and ${\C C}\!=\!\tilde{s}$, with the initial conditions $\tilde{s}(0)\!=\!0$ and $\tilde\chi(0)\!=\!1$. (c) The analytic prediction for the peak stress $\tilde{s}_p$ in Eq.~\eqref{eq:simplyfiedsp} (solid red line) compared to the peak stress obtained from a numerical solution of Eqs.~\eqref{eq:dimless_s}-\eqref{eq:dimless_chi} (open blue circles). (d) The prediction of the maximal stress drop rate $|\dot{\tilde{s}}_m|$ in the post-peak regime following Eq.~\eqref{eq:stress_drop} (solid red line) compared to the maximal stress drop rate obtained from a numerical solution of Eqs.~\eqref{eq:dimless_s}-\eqref{eq:dimless_chi} (open blue circles).}
\label{fig2}
\end{figure}

In Fig.~\ref{fig2}a-b we present $\tilde{s}(\tilde{t})$ and
$\tilde{\chi}(\tilde{t})$ for two values of $\xi$ which differ by an order of
magnitude. It is observed that as $\xi$ decreases, when $\tau^{ext}$ decreases
relative to $\tau^{pl}_0$, the yielding behavior of the material (i.e.~the
transition
from elastic-dominated to plastic-dominated deformation) changes quite
significantly. In particular, an elastic overshoot leads to a significant
increase in the peak stress $\tilde{s}_p$ with decreasing $\xi$, and the
subsequent dynamics exhibit a sharp drop in the stress $\tilde{s}$ and a sharp
increase in the effective temperature $\tilde\chi$. These sharp post-yielding
dynamics mark the emergence of a short timescale associated with strongly
nonlinear material response.

Our next goal is to better understand this
behavior and its relation to the fracture toughness.
To that aim, we first try to estimate the peak stress $\tilde{s}_p$, for which
$\dot{\tilde{s}}\!=\!0$. The latter translates into the relation
$e^{\tilde{e}_z(1-\tilde\chi_p^{-1})}(\tilde{s}_p-1)\!=\!\xi^{-1}$ between
$\tilde{s}_p$ and $\tilde\chi_p\!\equiv\!\tilde\chi(\tilde{s}_p)$.
An approximate solution for the stress peak can be derived in the form
\begin{equation}
\label{eq:simplyfiedsp}
\tilde{s}_p \simeq 1 -\frac{\zeta+\tilde\mu}{4 \zeta} +
\frac{\sqrt{8\,\zeta\,\tilde\mu\,\xi^{-1} +
(\zeta + \tilde\mu)^2}}{4 \zeta} \ ,
\end{equation}
with $\zeta\!\equiv\!1+\tilde{e}_z\left(\tilde{\chi}_\infty-1\right)$.
$\tilde\chi_p$ is given by the exact relation
$\tilde\chi_p(\tilde{s}_p)\!=\!\left(1+\tilde{e}_z^{-1}\log\left[\xi(\tilde{s}
_p-1)\right]\right)^{-1}$. In Fig.~\ref{fig2}c we compare the analytic
estimation in Eq.~\eqref{eq:simplyfiedsp} to the peak stress obtained from the
full numerical solution of Eqs.~\eqref{eq:dot_s}-\eqref{eq:dot_chi}. It is
observed that the analytic approximation accurately captures the increase in
$\tilde{s}_p$ with decreasing $\xi$. In light of the latter, we expect
$\tilde\chi_p(\tilde{s}_p)$ given above to yield good approximations as well,
which is indeed the case (not shown).

With $\tilde{s}_p$ and $\tilde\chi_p$ at hand, we can estimate the stress drop
rate in the post-peak dynamics, observed in Fig.~\ref{fig2}a. As the plastic
rate
of deformation is strongly amplified during the drop, we neglect the external
loading term $\dot{K}_I/\sqrt{2\pi\rho}$ in Eq.~\eqref{eq:dot_s}. With this
approximation, we can eliminate $D^{pl}(s,\chi)$ between
Eqs.~\eqref{eq:dot_s}-\eqref{eq:dot_chi}, obtaining a differential equation for
$\chi(s)$ (i.e.~time becomes a parameter). The solution, which is expected to be
valid deep inside the stress drop region, takes the form
$\tilde\chi(\tilde{s}
)\!\simeq\!\tilde\chi_\infty\!-\!(\tilde\chi_\infty\!-\!\tilde\chi_p)\exp\!{
\left(\!\frac{\tilde{s}^2-\tilde{s}_p^2}{2\tilde\mu}\!\right)}$. Using the latter, we
obtain the following estimate
\begin{equation}
\label{eq:stress_drop}
\dot{\tilde{s}}(\tilde{s})\!\simeq\!-\xi
e^{\tilde{e}_z\left[1-\tilde\chi(\tilde{s})^{-1}\right]}(\tilde{s}-1)
\end{equation}
for the stress rate during the drop, which should be valid for $1\!<\!\tilde{s}\!<\!\tilde{s}_p$, not too close to
either $1$ or $\tilde{s}_p$.

It is important to note that $\tilde\chi(\tilde{s})$ in
Eq.~\eqref{eq:stress_drop}
depends on $\xi$ also through $\tilde{s}_p$ and $\tilde\chi_p(\tilde{s}_p)$,
which give rise to a super-exponential increase in the maximal value of
$\dot{\tilde{s}}(\tilde{s})$, $|\dot{\tilde{s}}_m|$, with decreasing $\xi$. The
prediction in Eq.~\eqref{eq:stress_drop} is compared to the full solution in
Fig.~\ref{fig2}d, demonstrating reasonable agreement at small values of $\xi$. Note that
$|\dot{\tilde{s}}_m|$ in Fig.~\ref{fig2}d is one to two orders of magnitude
larger than the
effective external loading rate (which is unity in the dimensionless form, cf.
Eq.~\eqref{eq:dimless_s}) for
sufficiently small $\xi$, as assumed before. This analysis shows how nonlinear
yielding in glassy materials
can dynamically generate new, and much shorter, timescales.

\begin{figure}[h]
\centerline{\includegraphics[width=0.49\textwidth]{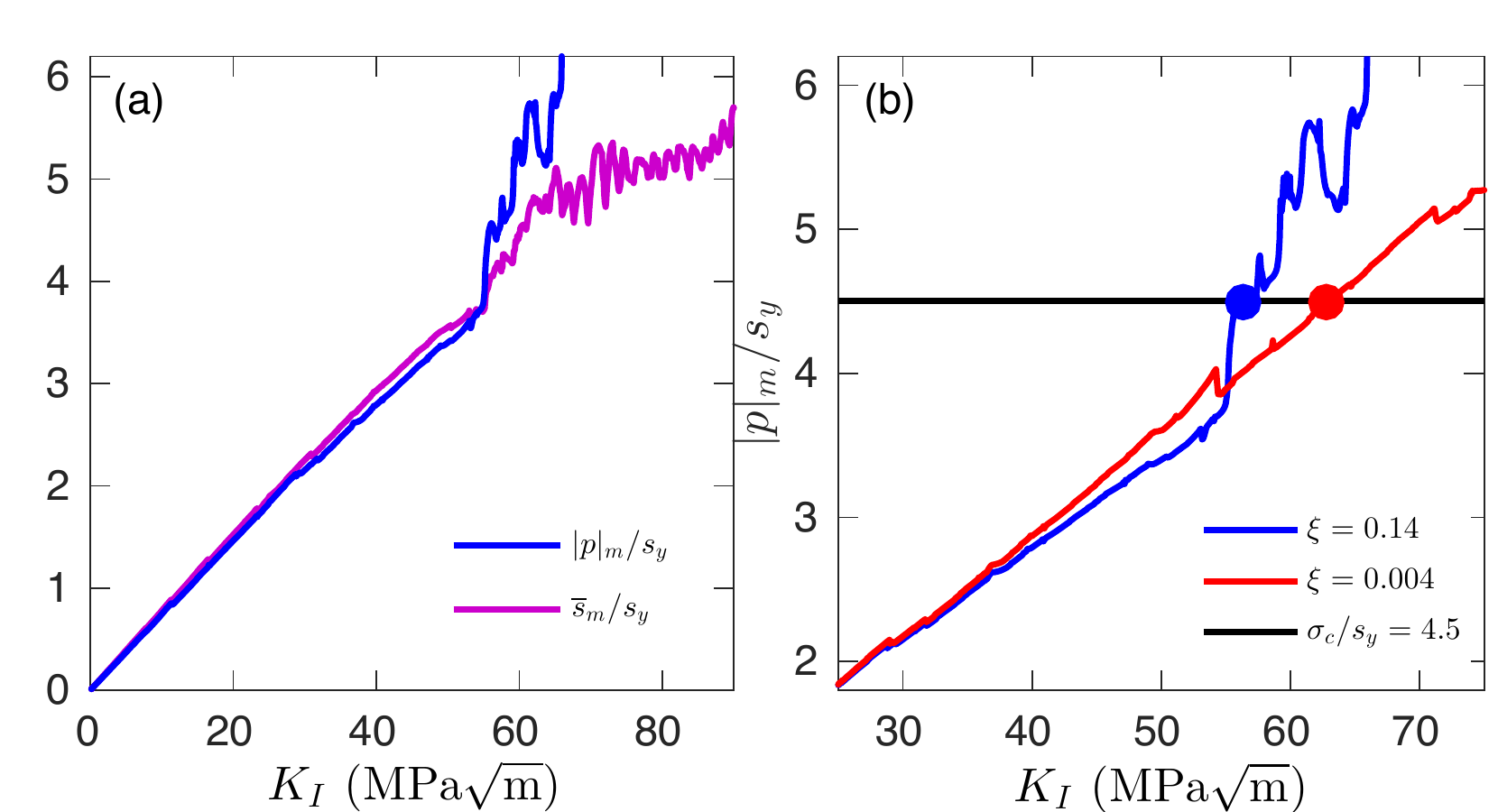}}
\caption{(Color online) (a) The maximum (in space) of the hydrostatic tension $|p|_m$ and the magnitude of the deviatoric stress $\bar{s}$ (both in units of $s_y$) as a function of $K_I$,
obtained from a numerical solution of the full 2+1 dimensional problem with $\chi_0\!=\!595~\rm{K}$, $\dot{K}_I\!=\!25~\rm{MPa\sqrt{m}/s}$ and $\rho\!=\!65~\mu\rm{m}$, corresponding to $\xi\!=\!0.14$. It is observed that the two quantities experience an abrupt increase at the same value of $K_I$. (b) The maximum (in space) of the hydrostatic tension $|p|_m$ (in units of $s_y$) as a function of $K_I$, for $\xi\!=\!0.14$ (as in panel (a)) and $\xi\!=\!4\!\times\!10^{-3}\!\ll\!1$, together with a cavitation threshold corresponding to $\sigma_c/s_y\!=\!4.5$ (solid black horizontal line). It is observed that for the larger $\xi$, the cavitation threshold is exceeded at a smaller $K_I$, implying a non-monotonic behavior of the fracture toughness.}
\label{fig3}
\end{figure}

In order to understand the implications of this reduced-dimensionality analysis
on the toughness, we need to consider spatial interactions between different
material elements and the coupling
between the deviatoric and the hydrostatic components of the stress tensor. When
a material element undergoes
yielding dynamics, the stress will be redistributed to nearby
material elements. In particular, when $\xi$ is small and a sharp deviatoric
stress drop as shown in Fig.~\ref{fig2}a emerges, nearby material elements will
experience a {\em sharp increase} in stress. This
applies to both the deviatoric $\B s$ and the hydrostatic
$\case{1}{3}\tr\B\sigma$ components of the stress tensor $\B \sigma$, which are
coupled through the stress equilibrium equation $\nabla\!\cdot\B\sigma\!=\!0$.
To show this, we plot in Fig.~\ref{fig3}a the maximum (in space) of the
magnitude of the deviatoric stress $\B s$, $\bar{s}_m$, and of the hydrostatic
tension
$\case{1}{3}\tr\B\sigma\!\equiv\!-p$, $|p|_m$, obtained from a numerical
solution of the full 2+1 dimensional problem with $\chi_0\!=\!595$~K and
$\dot{K}_I\!=\!25~\rm{MPa\sqrt{m}/s}$ (corresponding to
$\xi\!=\!0.14$).
We observe that indeed both quantities abruptly increase {\em together} at a
certain applied stress-intensity-factor
$K_I$.

If the increase in $|p|_m$ for the given $\xi$ is sufficiently large, it can
reach the threshold $\sigma_c$, which in our model implies failure (and hence the
toughness is determined). Consider then what happens for yet smaller values of
$\xi$, $\xi\!\ll\!1$, corresponding to larger $\dot{K}_I$'s or smaller
$\chi_0$'s. In this case, we expect the threshold $\sigma_c$ to be reached within the
predominantly elastic regime and hence
the toughness to be $\xi$-independent in this regime. This implies that there
might be a range of $\xi$'s in which the toughness decreases when $\xi$
increases. That is, this scenario implies that the toughness can
vary {\em non-monotonically} with $\xi$. To test this, we plot in
Fig.~\ref{fig3}b $|p|_m$ for $\xi\!=\!0.14$ (exactly the as in Fig.~\ref{fig3}a)
and also for $\xi\!=\!4\!\times\!10^{-3}\!\ll\!1$, along with $\sigma_c\!=\!4.5s_y$
(horizontal line, the same value as in~\cite{RB12}). We indeed observe that
the threshold is reached at a smaller $K_I$ for the larger $\xi$, i.e.~that the
fracture toughness is indeed non-monotonic.

\begin{figure*}
\centerline{\includegraphics[width=\textwidth]{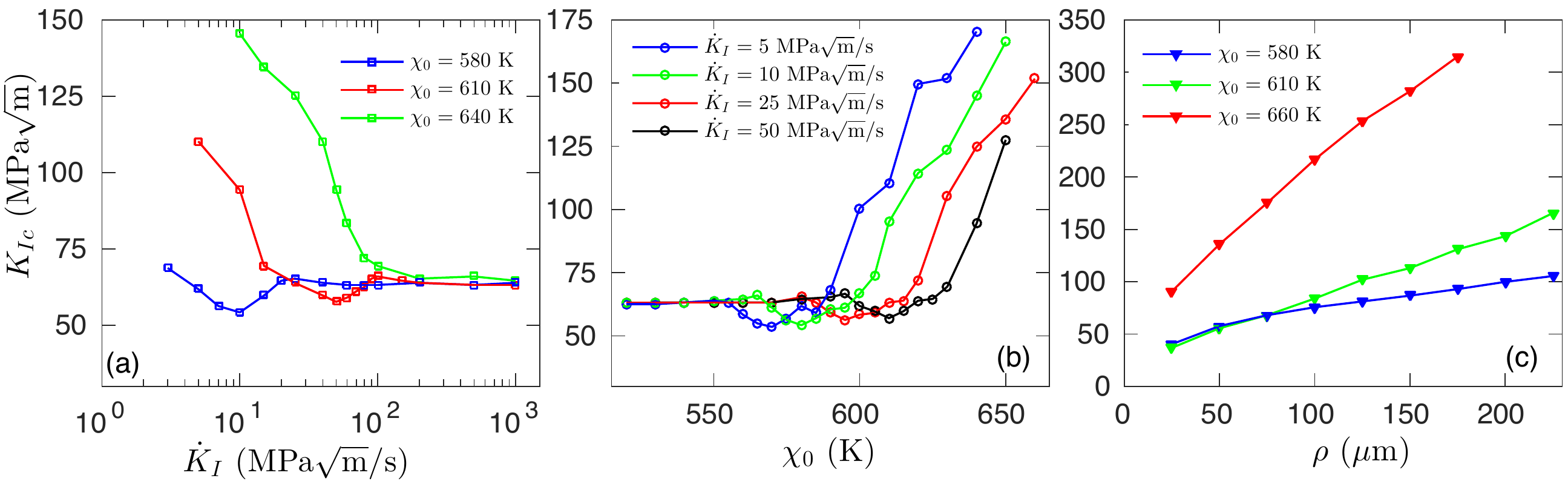}}
\caption{(Color online) The notch fracture toughness $K_{I\!c}(\chi_0, \dot{K}_I, \rho, T)$ as obtained from numerical solutions of the full 2+1 dimensional problem.
 (a) $K_{I\!c}$ as a function of $\dot{K}_I$ for various $\chi_0$'s, with $\rho\!=\!65~\mu\rm{m}$ and $T\!=\!400~\rm{K}$. (b) $K_{I\!c}$ as a function of the initial structural
state $\chi_0$ for various $\dot{K}_I$'s,  with $\rho\!=\!65~\mu\rm{m}$ and $T\!=\!400~\rm{K}$. (c) $K_{I\!c}$ as a function of
the notch radius $\rho$ for various $\chi_0$'s, with $\dot{K}_I\!=\!20~\rm{MPa\sqrt{m}/s}$ and $T\!=\!400~\rm{K}$.}
\label{fig4}
\end{figure*}

These predictions are tested over a wide range of
parameters in Fig.~\ref{fig4}a-b, where we plot the toughness as a function of
$\dot{K}_I$ (panel (a), for various $\chi_0$'s) and $\chi_0$ (panel (b), for
various $\dot{K}_I$'s), as obtained from the full 2+1 dimensional
computations. The emergence of a non-monotonic dependence of the toughness for
a large range of parameters is evident, as well as the
saturation of the toughness for sufficiently small $\chi_0$ and sufficiently
large $\dot{K}_I$ (corresponding to $\xi\!\ll\!1$). The minimum in the toughness
shifts systematically with $\chi_0$ and $\dot{K}_I$.
Note that while the non-monotonicity is not huge in magnitude, of the order of
$10$~MPa$\sqrt{\hbox{m}}$, it is a distinct and qualitative feature of strongly
nonlinear yielding dynamics in our model. Note also that the non-monotonic
behavior disappears for large enough $\chi_0$ (cf. the $\chi_0\!=\!640$~K curve
on panel (a)) and large enough $\dot{K}_I$ (not shown on panel (b), it requires yet
larger $\dot{K}_I$ values).

Finally, we also plot in Fig.~\ref{fig4}c the
variation of the toughness with $\rho$ (for various $\chi_0$'s, with $\dot{K}_I\!=\!20~\rm{MPa\sqrt{m}/s}$), to be discussed below. The toughness is
obviously a monotonically increasing function of $\rho$, as increasing the notch
radius of curvature implies enhanced plastic dissipation and less stress
concentration. Yet the monotonic $\rho$ dependence in panel (c) will be
connected below to the non-monotonic behavior observed in panels (a)-(b) with
respect to $\chi_0$ and $\dot{K}_I$.
\begin{figure}[h]
\centerline{\includegraphics[width=0.53\textwidth]{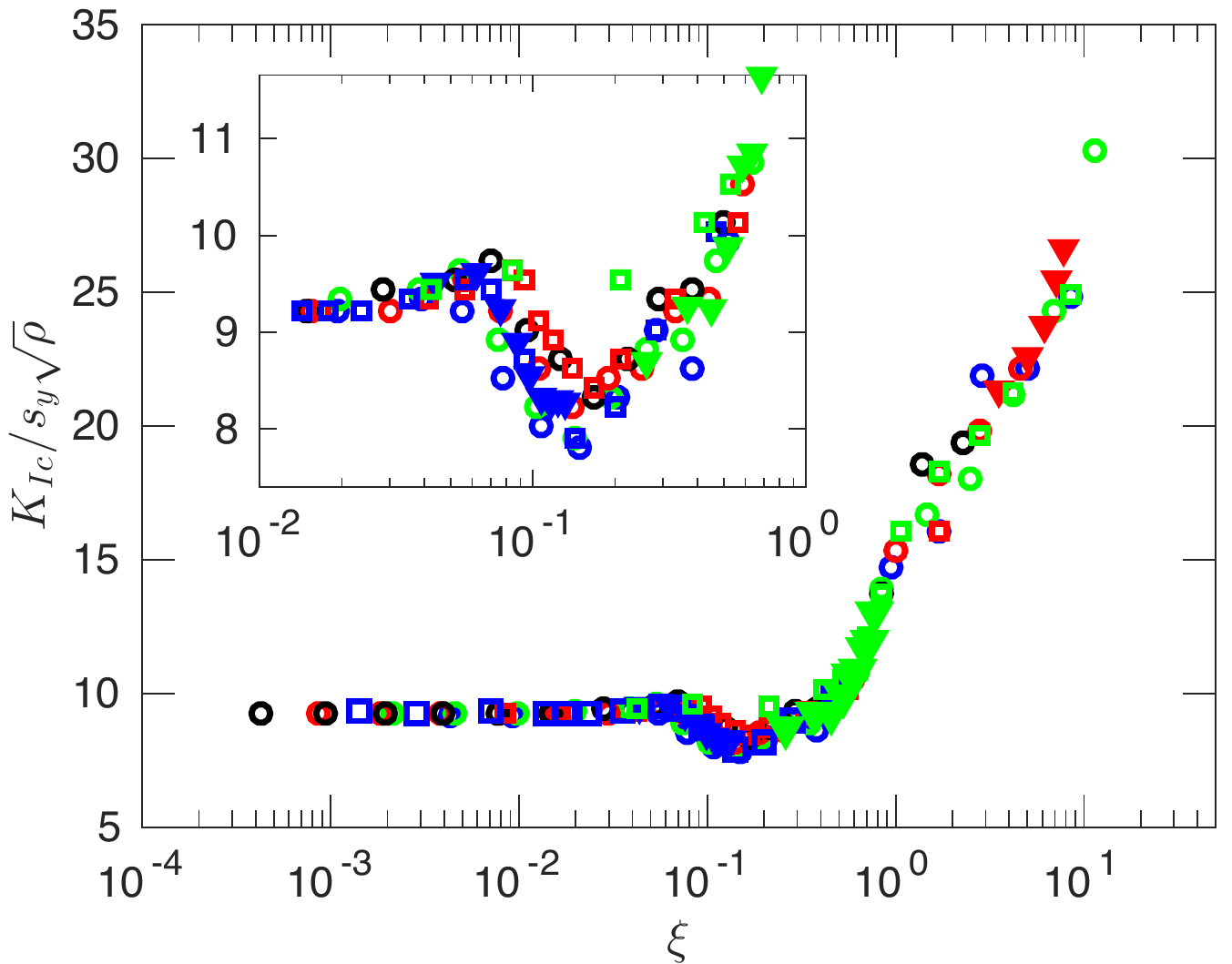}}
\caption{(Color online) The dimensionless notch toughness $K_{I\!c}(\chi_0, \dot{K}_I, \rho, T)/(s_y\sqrt{\rho})$ as a function of $\xi$, using all the data presented in Fig.~\ref{fig4}. As predicted theoretically in Eq.~\eqref{eq:toughness}, all data sets (except one in the non-monotonic part of the curve) collapse on a single master curve being a constant for $\xi\!\ll\!1$, varying as $\log(\xi)$ for $\xi\!\gg\!1$ and featuring a non-monotonic behavior for $\xi\!\sim\!{\cal O}(1)$. (inset) Zooming in on the non-monotonic part of the toughness master curve. Note in particular that data corresponding to the monotonic variation of the toughness with $\rho$ in Fig.~\ref{fig4}c nicely collapse on the non-monotonic part of the master curve (filled triangles) and that one data set (open green squares) do not exhibit a non-monotonic behavior (corresponding to the $\chi_0\!=\!640~\rm{K}$ data set in Fig.~\ref{fig4}a).}
\label{fig5}
\end{figure}
\section{The main result}
\label{sec:results}

Up to now, in the analysis of the reduced-dimensionality model in
Eqs.~\eqref{eq:dot_s}-\eqref{eq:dot_chi} we used ${\C C}\!\sim\!s$. This cannot
be valid in the large $\chi_0$ and small $\dot{K}_I$ limits (corresponding to
large $\xi$), where stresses remain close to $s_y$ and ${\C C}(\cdot)$ in
Eq.~\eqref{Cal_C} is expected to be determined by thermal activation.
Consequently, we would like now to gain some insight into the behavior of
Eqs.~\eqref{eq:dot_s}-\eqref{eq:dot_chi} when ${\C C}(s,T)\!=\! e^{-\Delta/k_B
T} \cosh\left[\Omega\,\epsilon_0\,s/k_B T\right]$. As the stress remains close
to $s_y$, we assume that $\chi\!\simeq\!\chi_0$ and expand $\dot{s}\!=\!0$ near
$s_y$. We can then solve for the peak stress, which takes the form
$\tilde{s}_p\!=\!1 + \psi^{-1} W(2\,\psi\,\xi^{-1} e^{\Delta/k_B T} e^{-\psi})$,
where $\psi\!\equiv\!\tfrac{\Omega\,\epsilon_0\,s_y}{k_B T}$ and $W(\cdot)$ is
the
Lambert W-function. For realistic numbers, the argument of the latter is large
and we have $W(x)\!\simeq\!\log(x)$. Consequently, $\tilde{s}_p$ depends on
$\xi$ through $T\log\xi$, a clear signature of thermal activation. We {\em
hypothesize} that the toughness $K_{I\!c}$ features the same dependence when
$\xi$ is large, i.e.~when stresses remain relatively small.

We are now ready to put all elements of the analysis into a unified prediction
for $K_{I\!c}(\chi_0, \dot{K}_I, \rho, T)$. The analysis above suggests that the
natural quantity to consider is actually $K_{I\!c}/\sqrt{\rho}$ (which can be
made dimensionless using a stress scale, say $s_y$). Consequently, we have
\begin{eqnarray}
\label{eq:toughness}
\hspace{-0.5cm}\frac{K_{I\!c}(\chi_0, \dot{K}_I, \rho, T)}{\sqrt{\rho}}  \sim
\left\{\begin{array}{lll}
\!\! \text{Const.} &\,\,\text{for}\quad\  \xi\!\ll\!1 \vspace{0.1cm}\\
\!\! g(\xi) &\,\,\text{for}\quad\  \xi\!\sim\!{\C O}(1) \vspace{0.1cm}\\
           \!\! \,\,T\,\log(\xi) &\,\,\text{for}\quad \ \xi\!\gg\!1  \ ,
\end{array}\right.
\end{eqnarray}
where $g(\xi)$ features a non-monotonic behavior for not too large $\chi_0$
and $\dot{K}_I$ (i.e.~$g(\xi)$ is not a unique function of $\xi$ for large enough
$\chi_0$
and $\dot{K}_I$). To test this major prediction, we re-plot in Fig.~\ref{fig5}
the data appearing
in Fig.~\ref{fig4} as $K_{I\!c}/s_y\sqrt{\rho}$ vs. $\xi$ in linear-log
scale. We observe that as predicted, all data collapse onto a single master
curve in the $\xi\!\ll\!1$ limit, where it is a constant, and in the
$\xi\!\gg\!1$ limit, where it varies as $\log(\xi)$, and feature a {\em
non-monotonic} behavior for $\xi\!\sim\!{\C O}(1)$ for a broad range of
parameters.

Note in particular the data corresponding to variations in $\rho$, which fall
onto the non-monotonic parts of the curve and on the $\log(\xi)$ part. That
means that while
$K_{I\!c}$ is monotonic in $\rho$, when the proper dimensionless variables are
used,
it can reveal the non-monotonic behavior of the toughness master curve.
Furthermore, it implies
that the dependence of $K_{I\!c}$ on $\rho$ differs from the existing
literature, both theoretical and experimental, where
$K_{I\!c}$ is expected to be either proportional to or linear
in $\sqrt{\rho}$, mainly based on dimensional arguments~\cite{Notch, FONYKI07, LSN06, HA09, XRM10, NTSNR15}. While this
dependence would give {\em apparently} reasonable fits to the data in
Fig.~\ref{fig4}c, our analysis shows that there
exists an {\em additional} and previously overlooked dependence on $\rho$
through $\xi(\chi_0, \dot{K}_I, \rho)$, for both $\xi\!\sim\!{\C O}(1)$ and
$\xi\!\gg\!1$.
\begin{figure*}[ht]
\centerline{\includegraphics[width=\textwidth]{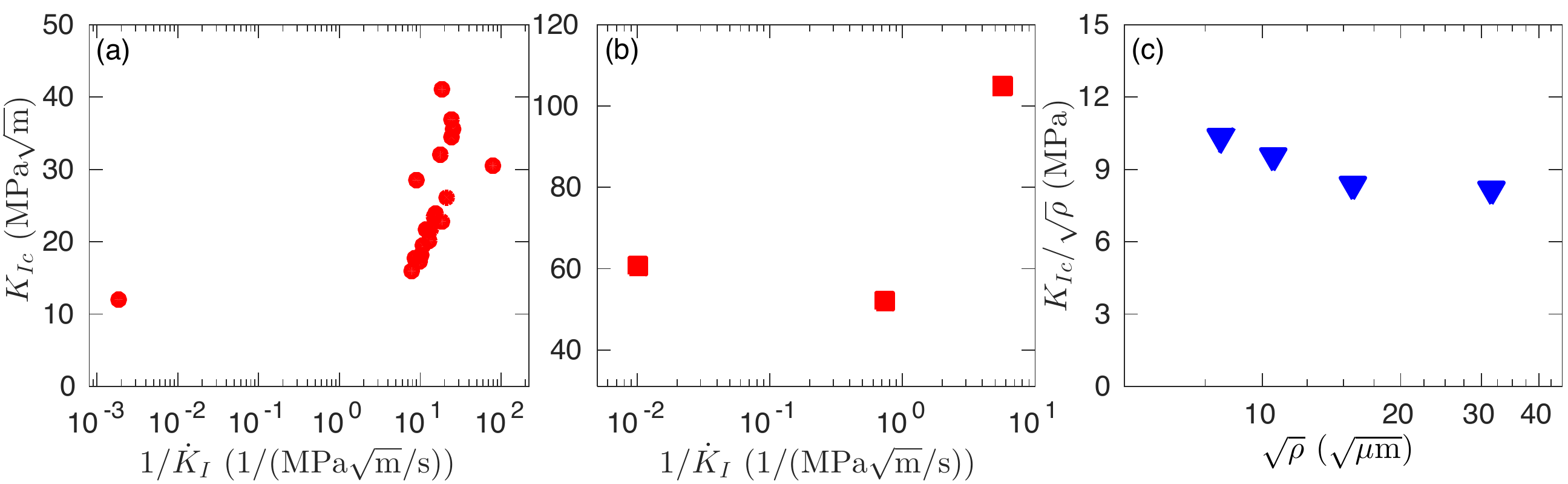}}
\caption{(Color online) Experimental support for Bulk Metallic Glasses (BMG). (a) The notch fracture toughness data $K_{I\!c}(\dot{K}_I)$ of \cite{FD01_1} re-plotted as $K_{I\!c}$ vs. $\log(1/\dot{K}_I)$, following the theoretical prediction in Eq.~\eqref{eq:toughness}. (b) The same as panel (a), but for the data of \cite{FONYKI07}. (c) The notch fracture toughness data $K_{I\!c}(\rho)$ of \cite{LSN06} re-plotted as $K_{Ic}/\sqrt{\rho}$ vs. $\log(\sqrt{\rho})$, following the theoretical prediction in Eq.~\eqref{eq:toughness}.}
\label{fig6}
\end{figure*}

It should be stressed that the crossover from the $\xi\!\ll\!1$ behavior to the
$\xi\!\gg\!1$ behavior, with the possible non-monotonicity, is directly related
to the change in the transition rate factor ${\C C}$ in Eq.~\eqref{Cal_C} with
increasing stress, from thermal activation at relatively small stresses to
athermal processes at higher stresses. This change in behavior, which is not
commonly discussed in the literature, implies that different parts of the
function $K_{I\!c}(\chi_0, \dot{K}_I, \rho, T)$ will depend differently on the
temperature (note, though, that we have not considered other possible
dependencies on the temperature). In particular, we have verified that the
$\log(\xi)$ dependence
originates from thermal activation (e.g. it disappears if thermal activation is
eliminated altogether and its logarithmic slope varies in
proportion to $T$), which suggests that glasses exhibit appreciable thermal
effects well below their glass temperature. These predictions can be tested by
systematically
performing notch toughness experiments at different temperatures.

Figure~\ref{fig5}, which summarizes our main result, provides a comprehensive
picture of the notch toughness of glasses and various testable predictions. It
shows that the toughness emerges from a competition between the {\em initial
(i.e.~far from steady-state)} plastic relaxation timescale, which depends on the
glass history/age, and an effective loading timescale near the notch root,
which depends on the global problem though $\dot{K}_I$ and on the notch geometry
through $\rho$. It also shows that the notch fracture toughness of glasses can
vary quite substantially, as was claimed in~\cite{RB12}, by changing $\xi$.
The toughness shown in Figs.~\ref{fig4}-\ref{fig5} implies a variation of more
than an order of magnitude in the fracture energy
$\Gamma\!\propto\!K^2_{I\!c}/\mu$~\cite{Meyers_Book}.

\section{Experimental evidence}
\label{sec:experiments}

While the toughness of glasses was experimentally studied by various groups,
there are relatively few works that systematically vary the
stress-intensity-factor rate, the age of the glass and the notch radius over a
large range. In
Fig.~\ref{fig6} we show three experimental data sets for BMG available in the
literature, where the notch toughness was measured as a function of $\dot{K}_I$
(panels (a) and (b)) and $\rho$ (panel (c)).

Inspired by the theoretical prediction in Eq.~\eqref{eq:toughness} and its
numerical validation in Fig.~\ref{fig5},
we re-plot in Fig.~\ref{fig6}a-b the data of~\cite{FD01_1} and of~\cite{FONYKI07},
respectively, as $K_{I\!c}$ vs. $\log(1/\dot{K}_I)$.
The data in panel (a) are consistent with our predictions as they feature a
quasi-logarithmic dependence on
$\log(1/\dot{K}_I)$ for small $\dot{K}_I$ and indicate the existence of a
plateau for large $\dot{K}_I$'s. There is, however, a gap of nearly $4$ orders
in magnitude in $\dot{K}_I$ in the data,
so the possible non-monotonic behavior at intermediate $\dot{K}_I$'s cannot be
tested. The data on panel (b) feature all of the predicted trends, including a
non-monotonicity of a similar magnitude compared to our prediction, though there
are too few experimental points to test functional dependencies.

In Fig.~\ref{fig6}c we re-plot the data of~\cite{LSN06} as $K_{I\!c}/\sqrt{\rho}$ vs.
$\log(\sqrt{\rho})$.
The experimental data, where $\rho$ ranges from $65~\mu$m to $250~\mu$m, seem to
be consistent with
the decreasing part of the toughness master curve in Fig.~\ref{fig5} and
possibly indicate the existence of a minimum.
A broader range of $\rho$'s, together with varying also $\chi_0$ and $\dot{K}_I$, are needed in order to test the predicted functional
dependencies, but it is already clear that re-plotting the existing data
inspired by to our theory reveals new features of the toughness. Our theory
certainly calls for additional experiments, as it offers various new qualitative
and quantitative predictions.

\section{Concluding remarks and prospects}
\label{sec:conclusions}

In this paper we provided a comprehensive theory of the notch fracture toughness
of glassy solids,
focussing on its dependence on the structural state of the glass (quantified by
the initial value of the effective disorder temperature $\chi_0$),
on the stress-intensity-factor rate $\dot{K}_I$, on the notch radius of
curvature $\rho$ and on the temperature
$T$ below the glass transition temperature. The main results are the theoretical
prediction in Eq.~\eqref{eq:toughness} and its numerical validation in
Fig.~\ref{fig5}
based on a novel computational method~\cite{RSB15}. The theory highlights the
underlying competition between an intrinsic plastic relaxation timescale and an
extrinsic driving timescale, as well as the roles played by nonlinear yielding
dynamics and a crossover between thermal/athermal rheological processes.
The theoretical predictions are shown to be consistent with existing experimental data.

The analysis presented has been based on a simple version of the non-equilibrium
thermodynamic Shear-Transformation-Zones (STZ) model~\cite{RB12}. We suspect that
despite its relative simplicity, the model captures some salient features of
glassy rheology, accounting for generic properties
of the notch fracture toughness of glasses. More elaborate models and
quantitative predictions will be explored in the future. Other physical effects
that were already identified in our numerical solutions, such as the time
evolution of the notch curvature $\rho(t)$ and the propagation of plastic
yielding fronts, will
be reported on in a subsequent publication, along with discussing the
post-cavitation dynamics. The latter were shown in~\cite{RB12} to lead to
catastrophic failure, as we assumed in this work, though we did not discuss them
at all.

A few important directions for future investigations emerge from the present
analysis. Most notably, one would be interested in calculating the {\em
intrinsic}
toughness, as opposed to the notch toughness, in the limit of $\rho\!\to\!0$,
where the notch/tip radius of curvature is not the dominant lengthscale in the
problem. This touches upon a fundamental problem in glass physics, i.e.~the
existence on an intrinsic glassy lengthscale. Within the adopted non-equilibrium
thermodynamic framework, such a lengthscale may appear in the macroscopic theory
in an effective diffusion term proportional to $\nabla^2\chi$ in
Eq.~\eqref{eq:chi}~\cite{MLC07, Kamrin2014}. This will be discussed in separate report. Finally,
it would be
interesting to see whether variations in the glass composition, and their effect
of the toughness, can be incorporated into the proposed theoretical framework.

\textit{Acknowledgments} E.B.~acknowledges support from the Israel Science
Foundation (Grant No.~712/12), the Harold Perlman Family Foundation and the
William Z. and Eda Bess Novick Young Scientist Fund. C.H.R. was supported by the National Science Foundation under Grant No. DMR-1409560, and by the Director, Office of Science, Computational and Technology Research, U.S. Department of Energy under contract number DE-AC02-05CH11231.

\end{document}